\documentclass{article}
\usepackage[english]{babel}
\usepackage[letterpaper,top=2cm,bottom=2cm,left=3cm,right=3cm,marginparwidth=1.75cm]{geometry}
\usepackage{amsmath,amssymb,amsfonts,amsthm,authblk,mathtools,empheq,cite}
\usepackage{graphicx}
\usepackage{multirow}
\usepackage[colorlinks=true, allcolors=blue]{hyperref}
\newcommand{\p}{\partial}
\newcommand{\R}{\mathbb{R}}

\newtheorem{remark}{Remark}[section]
\newcommand*\diff{\mathop{}\!\mathrm{d}}
\newtheorem*{remark*}{Remark}
\usepackage[capitalise]{cleveref}

\usepackage{hyperref}
\usepackage{algorithm}
\usepackage[noend]{algpseudocode}
\usepackage[mathlines]{lineno}

\title{Learning Structured Population Models
from Data with WSINDy}
\author[1]{Rainey Lyons\footnote{rainey.lyons@colorado.edu}}
\author[1]{Vanja Dukic}
\author[1]{David M.~Bortz}

\affil[1]{Department of Applied Mathematics, University of Colorado, Boulder CO 80309-0526, USA}

\begin{document}

\maketitle

\begin{abstract}

In the context of population dynamics, identifying effective model features, such as fecundity and mortality rates, is generally a complex and computationally intensive process, especially when the dynamics are heterogeneous across the population.
In this work, we propose a Weak form Scientific Machine Learning-based method for selecting appropriate model ingredients from a library of scientifically feasible functions used to model structured populations.
This method uses extensions of the Weak form Sparse Identification of Nonlinear Dynamics (WSINDy) method to select the best-fitting ingredients from noisy time-series histogram data.
This extension includes learning heterogeneous dynamics and also learning the boundary process of the model directly from the data.
We additionally provide a cross-validation method which helps fine tune the recovered boundary process to the data.

Several test cases are considered, demonstrating the method's performance for different previously studied models, including age and size-structured models. 
Through these examples, we examine both the advantages and limitations of the method, with a particular focus on the distinguishability of terms in the library.

\end{abstract}
\section{Introduction}

The study of structured population dynamics is an active field in mathematical biology where the resulting models and theory provide a rigorous framework for describing how individual-level traits, such as age, size, or physiological state, influence the  dynamics of a population.
Among these models, hyperbolic partial differential equations (PDEs) have proven particularly effective in contexts where one can model individuals' state progression as a deterministic transport process. 
This class of models arises naturally in applications ranging from cell growth and organism development \cite{MetzDiekmann1986a,VonFoerster1959TheKineticsofCellularProliferation} to disease progression and epidemiology \cite{KeyfitzKeyfitz1997MathematicalandComputerModelling,McKendrick1925ProcEdinbMathSoc}.

Two of the most well-studied models in this setting are the age-structured model, also known as the McKendrick–von Foerster model  \cite{McKendrick1925ProcEdinbMathSoc,VonFoerster1959TheKineticsofCellularProliferation}, and the size-structured model, or Sinko–Streifer model  \cite{SinkoStreifer1967Ecologya}. 
These equations exemplify the typical hyperbolic structure of these models: transport and reaction processes, which make use of \emph{growth}, \emph{death}, and \emph{birth} functions to encode both internal and external regulatory mechanisms. 
For instance, an age-structured model describes the evolution of the population number density, $n$, using the system
\begin{subequations}\label{Eq:AgeStructured}
   \begin{empheq}[left=\empheqlbrace]{align}
       &\p_t n + \alpha \p_a n = -d[n](a)n, & (t,a) \in (0,T) \times (0,\infty)\label{Eq:AgeStructuredPDE}\\
       &n(t,0) = \int_0^\infty \beta[n](a)n(t,a) \diff{a},& t \in (0,T), \label{Eq:AgeStructuredBC}\\
       &n(0,a) = n_0(a),& a \in  \R^+,\label{Eq:AgeStructuredIC}
   \end{empheq} 
\end{subequations}
where $a\in\R^+:=[0,\infty)$ represents the age of individuals,
and a size-structured version is given by
\begin{subequations}\label{Eq:SizeStructured}
   \begin{empheq}[left=\empheqlbrace]{align}
       &\p_t n +\p_x(g[n](x) n) = -d[n](x) n, & (t,x) \in (0,T)\times (x_{1},x_{2}),\label{Eq:SizeStructuredPDE}\\
       &g[n](x_{1}) \, n(t,x_{1}) = \int_{x_{1}}^{x_{2}} \beta[n](x)n(t,x) \diff{x}, & t \in (0,T),\label{Eq:SizeStructuredBC}\\
       &g[n](x_{2}) \, n(t,x_{2}) = 0, & t \in (0,T),\label{Eq:SizeStructuredBC2}\\
       &n(0,x) = n_0(x), & x \in [x_{1},x_{2}],\label{Eq:SizeStructuredIC}
   \end{empheq} 
\end{subequations}
where $x \in [x_{1},x_{2}] \subset \R^+$ represents the size of individuals.
Here, we use the notation $f[n]$ to denote a non-pointwise dependency on the population density which is usually nonlocal in nature. 
Details on the specific structure are provided in the next section.
In both the age structure and size structured setting, the accuracy and interpretability of the model  depend heavily on the functional forms and parameters of the growth, death, and birth terms, denoted in the above equations by $g$, $d$, and $\beta$, respectively.
Generally, these forms can be inferred by studies at the individual level, but in many instances such studies are expensive or infeasible, and one may not be able to obtain an effective dependence of the model ingredients on the structural variable \cite{MilnerRabbiolo1992JMathBiol,PilantRundell1991SIAMJApplMatha}.
Additionally, these vital components may depend directly on environmental variables such as the total population \cite{GurtinMaccamy1974ArchRationalMechAnal} or resource abundance \cite{Cushing1990RockyMountainJMatha}, which tend to be processes that cannot be directly measured and whose functional form is unknown and is usually taken as an \textit{ansatz}.

Classical approaches to model calibration rely on specifying parametric forms for these functions and estimating parameters by minimizing the discrepancy between simulated and observed population dynamics (see, for instance, \cite{Wood1997Structured-PopulationModelsinMarineTerrestrialandFreshwaterSystemsa} and the references therein).
While effective, this approach is computationally demanding, as it requires repeated forward simulations of the PDE system. 
Furthermore, parameter estimation problems are often ill-posed, especially in the presence of noise, limited data, or structural uncertainty in the model.
Recently, Weak form Scientific Machine Learning (WSciML) methods such as the Weak form Sparse Identification of Nonlinear Dynamics (WSINDy)\footnote{A WSciML extension of the well known Sparse Identification of Nonlinear Dynamics \cite{BruntonProctorKutz2016ProcNatlAcadSci,RudyBruntonProctorEtAl2017SciAdv} for equation discovery.} \cite{MessengerBortz2021JComputPhys,MessengerBortz2021MultiscaleModelSimul}, and the the Weak-form Estimation of Nonlinear Dynamics (WENDy) \cite{BortzMessengerDukic2023BullMathBiol,RummelMessengerBeckerEtAl2025arXiv250208881} have emerged as promising alternatives for learning governing equations and estimating parameters directly from data. 
These methods bypass the need for repeated forward simulations by instead minimizing an equation error residual over all possible combinations  of candidate terms in the library. 
In particular, the weak-form algorithms, WSINDy and WENDy, have been shown to be robust to noise, while retaining high accuracy and computational efficiency (for a general overview of weak-form methods see e.g., \cite{BortzMessengerTran2024NumericalAnalysisMeetsMachineLearning,MessengerTranDukicEtAl2024SIAMNews}).

In this work, we apply the WSINDy framework to the discovery of hyperbolic structured population equations.
Given noisy population data, our goal is to identify effective model components from a library of biologically plausible functions. 
From the list of aforementioned methods, WSINDy is the most natural choice of method for the considered problem, as structured population models are commonly studied in a weak sense (as smoothness is not always guaranteed \cite{AcklehIto2005JDifferEqu,AcklehLyonsSaintier2021ESAIMM2AN,DullGwiazdaMarciniak-CzochraEtAl2021}).
We demonstrate the performance of this approach on both synthetic and real data, showing that WSINDy recovers relevant and effective dynamics with significantly reduced computational effort compared to traditional parameter estimation methods.

Finally, we highlight several novel aspects of our work. 
Previous implementations of the WSINDy algorithm have focused primarily on discovering homogeneous, pointwise nonlinearities of the form $f(n)$.
Here, we explore the method's ability to identify both heterogeneous model ingredients and boundary processes directly from the data.
To our knowledge, both of these extensions are yet unexplored in the context of weak form model selection.
Additionally, we consider nonlocal nonlinearities of the form $f(s, N(t))$, which are ubiquitous in structured population models and capture well-known density-dependent effects such as those found in logistic-type nonlinearities, Ricker suppression terms, and Beverton-Holt-type population effects \cite{Getz1980MathematicalBiosciences,GurtinMaccamy1974ArchRationalMechAnal,Perthame2007}.  
These features introduce new challenges that we address in the sections that follow.

The manuscript is organized as follows: in \cref{Sec:ModelsandMethods}, we present the modified WSINDy method for a general structured population model and include details in how to incorporate the identification of the boundary process. 
In \cref{Sec:Results}, we discuss the assumptions made on the data and present an array of test problems which are focused on the two most popular structured population models \eqref{Eq:AgeStructured} and \eqref{Eq:SizeStructured}.
In \cref{Sec:Noise}, we explore the performance of the method, including how the method performs in high noise cases and how well the method can distinguish between similar terms in the library.
Finally, in \cref{Sec:Discussion} we conclude with a discussion of the method and directions for future exploration.


\section{Models and Methods}\label{Sec:ModelsandMethods}
In this section, we present the general framework of the manuscript.
We begin by introducing a general structured population model which encompasses the well-studied models given by \cref{Eq:AgeStructured} and \cref{Eq:SizeStructured} as well as other commonly seen structured population models.
We then present the method applied to this general model and describe a validation procedure used to improve the recovery of the boundary terms. 

\subsection{A General Structured Population Model}\label{SubSec:GeneralSetting}
Throughout the manuscript, we assume that the noise-free data follows underlying dynamics governed by a hyperbolic structured population equation. 
To distinguish between noisy and clean data, we make use of the superscript $^\star$ to denote the noise-free population density and true model ingredients. 
In such models, the population is distributed over some structural variable (such as age, size, or a combination with other physiological traits), and its distribution is denoted by $n^\star(t,s)$, where $t \in [0,T]$ denotes time and $s \in \Omega \subseteq \R^{d}$ denotes an arbitrary structural variable, assumed to lie in a connected domain $\Omega$ with boundary $\partial \Omega = \partial\Omega^+ \cup \partial\Omega^-$. 
Another natural interpretation of this density $n^\star$  commonly found in ecology is that the number of individuals whose structure lies in a measurable subset $S \subset \Omega$ at time $t$ is given by the quantity $\int_S n^\star(t,s) \diff{s}$.  (Note that the number density integrates to the total population size, $N(t) = \int_\Omega n(t,s) \diff{s}$.)
While the rest of this work focuses primarily on the most common case $d = 1$, we note that there is no inherent restriction of the method to one-dimensional models.

The following system gives a general form of the dynamics:
\begin{subequations}\label{Eq:GeneralStructured}
   \begin{empheq}[left=\empheqlbrace]{align}
       &\p_t n^\star(t,s) +\nabla_s\cdot(g^\star[n^\star](s) \, n^\star(t,s)) = f^\star[n^\star](s,n^\star), &\quad (t,s) \in (0,T) \times \Omega, \label{Eq:GenStructuredPDE}\\
       &g^\star[n^\star](s)\, n^\star(t,s) \cdot \vec{\eta}(s) = \int_\Omega \beta^\star[n^\star](\sigma) n^\star(t,\sigma) \diff{\sigma}, &\quad (t,s) \in [0,T]\times \p \Omega^+ ,\label{Eq:GenStructuredBC1}\\
       &g^\star[n^\star](s)\, n^\star(t,s) \cdot \vec{\eta}(s) = 0, &\quad (t,s) \in [0,T]\times \p \Omega^- ,\label{Eq:GenStructuredBC2}\\
       &n^\star(0,s) = n^\star_0(s),& \quad s \in \Omega. \label{Eq:GenStructuredIC}
   \end{empheq} 
\end{subequations}
In the equations above, $\vec{\eta}(s)$ denotes the inward-pointing unit normal vector at the boundary point $s$. 
The model terms $g^\star$, $f^\star$, and $\beta^\star$ represent biological processes which modeled via transport, source, and boundary terms. 
Common examples of such processes are growth/aging, mortality or division, and reproduction, respectively. 
These terms will be assumed to be smooth or at least globally Lipschitz in all arguments to be consistent with the well-posedness theory for \cref{Eq:GeneralStructured} (see, e.g., \cite{DullGwiazdaMarciniak-CzochraEtAl2021}). 
The notation $f[n]$ denotes a non-pointwise dependence on the population density, typically through a weighted average of the population. 
For instance, $f[n](\cdot) = f(\cdot,\int_\Omega \gamma(s) n(t,s) \diff{s})$ for some known weight function $\gamma$. 
While many types of kernels are biologically relevant, well-studied, and have many interesting mathematical properties \cite{FalsterBrannstromDieckmannEtAl2011JournalofEcology,KooijmanMetz1984EcotoxicologyandEnvironmentalSafety,AcklehIto2005JDifferEqu}; in this manuscript, we focus on the most common biologically relevant nonlinearity (where $\gamma \equiv 1$), meaning that the model ingredients depend on the total population size $N(t) = \int_\Omega n(t,s) \diff{s}$.   

As one of the crucial components of the WSINDy method is the eponymous \emph{weak form}, integration over both the temporal and structural variables will play a vital role. 
To streamline notation, we define the $L^2$ inner product over time and space by
\[
\langle f, g \rangle := \int_0^T \int_\Omega f(t,s) \cdot g(t,s) \diff{s} \diff{t}.
\]
We will also abuse this notation slightly to represent discrete approximations of this inner product in the same way.

The weak form of \cref{Eq:GeneralStructured} is then given by:
\begin{equation}\label{Eq:GeneralWeakForm}
    -\langle \p_t \phi , n^\star \rangle - \langle \nabla_s \phi , g^\star[n^\star] n^\star \rangle = \langle \phi, f^\star[n^\star](\cdot,n^\star) \rangle,
\end{equation}
where $\phi(t,s)$ is a smooth real-valued function compactly supported in $(0,T) \times \Omega$, i.e, $\phi \in C^p_c((0,T) \times \Omega)$ for some $p \geq 2$. 
The test function $\phi$ plays a role similar to a Gaussian smoother, while also maintaining the quantitative relationship given by \cref{Eq:GeneralWeakForm}, allowing us to simultaneously smooth the data and utilize equation error methods.
The smoothness and compact support of the test function allows us to exploit the rapid convergence of the trapezoidal rule \cite[Lemma 2]{MessengerBortz2021MultiscaleModelSimul} allowing for highly accurate computations of the integrals appearing in the weak form.
However, one down side of these test functions is that the boundary conditions \eqref{Eq:GenStructuredBC1}-\eqref{Eq:GenStructuredBC2} and the initial condition \eqref{Eq:GenStructuredIC} are absent from \cref{Eq:GeneralWeakForm}.
This presents a challenge when attempting to identify the boundary process (commonly representing birth), a critical component in describing the population dynamics.
To address this, we provide a method for identifying the boundary process in the following section.


\subsection{Weak Sparse Identification of Nonlinear Dynamics (WSINDy)}\label{Sec:WSINDy}

In this section, we extend the WSINDy algorithm to accommodate structured populations. 
This extension is novel as it enables the identification of heterogeneous dynamics and boundary processes directly from the given data. 
In the subsequent section, we introduce a cross-validation procedure designed to leverage the accuracy of the learned boundary processes for hyperparameter tuning.

Let $\textbf{n}:=\{n_{j}\}_{j = 1}^{J}$ denote (after a suitable indexing) a set of possibly noisy observations of the  number density over a disjoint partition $\{\Lambda_j \}_{j = 1}^J \subset \Omega$, i.e., $n_j := \frac{1}{|\Lambda_j|}\int_{\Lambda_j}n(t_j,s) \diff{s}$.
As stated before, we assume the noise-free continuous density $n^\star$ evolves according to a true model of the form \eqref{Eq:GeneralStructured}. 
The WSINDy algorithm utilizes the weak form \eqref{Eq:GeneralWeakForm} to construct a sparse regression problem, selecting the proper model ingredients from a set of given trial functions, known as the ``library''. 
More precisely, we assume the true model ingredients $g^\star$, $f^\star$, and $\beta^\star$ can be represented as a (sparse) linear combination of a given set of trial functions.
The set of these functions is denoted by $\{g_m(s,N) \}_{m=1}^{M^g}$, $\{f_m(s,n,N) \}_{m=1}^{M^f}$, and $\{\beta_m(s,n,N) \}_{m=1}^{M^\beta}$, respectively.
Then, for a given set of test functions $\{ \phi_k\}_{k=1}^K \subset C^p_c((0,T)\times \Omega)$, we construct the linear system 
\begin{equation}\label{Eq:LinSystemPDE}
    \textbf{b}^{\text{pde}} = G^{\text{pde}} \textbf{w} = (G^g | G^f)\left( \frac{\ \textbf{w}^g\ }{\ \textbf{w}^f\ }\right),
\end{equation}
 where the vector $\textbf{b}^{\text{pde}} \in \R^{K \times 1}$  and matrix $G^{\text{pde}} \in \R^{K\times(M^g+M^f)}$ are given by 
\[\textbf{b}_k^{\text{pde}} :=-\langle\p_t \phi_k, n \rangle, \quad G^g_{k,m} := \langle \nabla_s \phi_k, g_m(\cdot,N)n  \rangle ,
\text{ and } G^f_{k,m} := \langle \phi_k, f_m(\cdot,n,N) \rangle.\]

As for the choice of test functions, we opt to make use of piecewise-polynomial test functions of the form $\phi(t,s) = \varphi(t) \prod_{i = 1}^{\diff{}} \psi_i(s_i)$, where
\[ 
\varphi(t) = 
\begin{cases}
    C_t (t - t_1)^p (t_2 - t)^q, &t \in (t_1 , t_2)\\
    0& \text{otherwise}
\end{cases} 
\] 
\text{and}
\[ 
\quad 
\psi_i(s_i) = 
\begin{cases}
    C_{i} (s_i - a_i)^p (s_2 - b_i)^q, & s_i \in (a_1 , b_2)\\
    0& \text{otherwise}
\end{cases}
\]
The constants $C_t$ and the $C_i$ are chosen such that $\|\varphi\|_\infty = \|\psi_i\|_\infty = 1$.
The supports of the test functions are determined from the data in such a way that the intervals $[a_i,b_i]$ or $[t_1,t_2]$ account for a given percentage of the full domain (denoted by $r_s$ and $r_t$, respectively).
For the numerical results to follow, we will set $p = q = 14$ and $r_s = r_t = 0.5$.
These choices were determined through numerical experiments, and properly choosing these parameters from the data is currently a frontier of active research.
We point out that while there exist methods for choosing the support and smoothness of test functions from the given data (such as, e.g., \cite{MessengerBortz2021JComputPhys}), these methods are not tuned for heterogeneous libraries such as those considered here.
Additionally, while this class of test functions is commonly used, we make no claims that it is indeed the best choice. 

As discussed in the previous section, since this choice of test function is compactly supported in $\Omega$, the boundary condition \eqref{Eq:GenStructuredBC1} is absent from the weak form \eqref{Eq:GeneralWeakForm}.
One natural way to account for the boundary is to couple \cref{Eq:GeneralStructured} with the dynamics of the total population $N(t) := \int_{\Omega} n(t,s) \diff{s}$ given by the ordinary differential equation acquired by integrating \cref{Eq:GeneralStructured}:
\begin{equation}\label{Eq:GenTotalPopDE}
    \frac{\diff{}}{\diff{t}}N = \int_{\Omega}\beta^\star[n](s) n(t,s)\diff{s} + \int_{\Omega} f^\star[n](s,n(t,s)) \diff{s},
\end{equation}
with the initial condition $N(0) = \int_\Omega n_0(s) \diff{s}$.
This equation then has the corresponding weak form
\begin{equation}\label{Eq:GenTotalPopWF}
    -\int_0^T \frac{\diff{}}{\diff{t}}\varphi(t) \, N(t) \diff{t} = \langle \varphi , \beta^\star[n] n\rangle + \langle\varphi , f^\star[n](\cdot,n) \rangle,
\end{equation}
for a smooth test function $\varphi$ with compact support in $(0,T)$. 

This allows us to extend or ``stack" the linear system \eqref{Eq:LinSystemPDE} by concatenating it with the linear system
\begin{equation}\label{Eq:LinSystemODE}
    \textbf{b}^{\text{ode}} = \Xi \textbf{v} : = \left(\Xi^f | \Xi^\beta \right) \left( \frac{\ \textbf{w}^f\ }{\ \textbf{w}^\beta\ }\right),
\end{equation}
with $\textbf{b}_k^{\text{ode}} \in \R^{K \times 1}$ and $\Xi \in \R^{K\times (M^f+M^b)}$ given by
\[\textbf{b}_k^{\text{ode}} := -\int_0^T \frac{\diff{}}{\diff{t}} \varphi_k(t) N(t) \diff{t} ,\quad \Xi_{k,m}^f :=   \langle\varphi_k, f_m(\cdot, n,N) \rangle, \quad \text{and} \quad \Xi_{k,m}^\beta := \langle \varphi_k, \beta_m(\cdot,N) n \rangle. \]
We then concatenate the two systems, resulting, finally, in the linear system 
\begin{equation}\label{Eq:FullWFSystem}
    \left(\begin{array}{c}
    \textbf{b}^{\text{pde}} \\
    \hline
     \textbf{b}^{\text{ode}}
\end{array} \right) =:\textbf{b} = G \textbf{w} := \left(\begin{array}{c|c|c}
    G^g & G^f & \textbf{0} \\
    \hline
    \textbf{0} & \Xi^f & \Xi^\beta
\end{array}\right)
\left(\begin{array}{c}
    \textbf{w}^g \\
    \hline
    \textbf{w}^f \\
    \hline
    \textbf{w}^\beta 
\end{array}\right).
\end{equation}

The problem then becomes finding sparse $\textbf{w}  \in \R^{(M^g + M^f+M^\beta) \times 1}$ which minimizes the loss function 
\begin{equation}\label{Eq:WSINDy_LossFunc}
    \mathcal{L}(\textbf{x}; \textbf{b},G, \lambda) := \|\textbf{b} - G \textbf{x}\|_2 + \lambda \|\textbf{x}\|_0, 
\end{equation}
where $\lambda$ is a given sparsity parameter. 
The function $\mathcal{L}$ is minimized using the modified sequential-thresholding least-squares method (MSTLS) provided in \cite{MessengerBortz2021JComputPhys}, which has been successful in a variety of applications of sparse regression techniques.
This method has been studied for differential equation models of various types and contexts, including ordinary differential equations \cite{MessengerBortz2021MultiscaleModelSimul}, partial differential equations \cite{MessengerBortz2021JComputPhys,MinorMessengerDukicEtAl2025arXiv250100738}, and hybrid models with multiple time scales \cite{MessengerDwyerDukic2024JRSocInterface}.

\begin{remark}
When studying population dynamics from an age-structured perspective, it is quite common to know \textit{a priori} the age-time relationship $\alpha$ in \cref{Eq:AgeStructured}.
In that case, one does not need to identify the transport term and can focus all efforts on the identification of the death and birth functions.
One can easily modify the method presented above by simply adding the transport term into the vector $\textbf{b}^{\text{pde}}$ in \cref{Eq:LinSystemPDE}. 
That is, \cref{Eq:LinSystemPDE} would be given by
\[\textbf{b}_k^{\text{pde}} :=-\langle\p_t \phi_k, n \rangle - \langle \nabla_s \phi_k, \alpha n  \rangle
\text{ and } G_{k,m}^{\text{pde}} := \langle \phi_k, f_m(\cdot,n,N) \rangle.\] 
\end{remark}

\subsubsection{Boundary Bagging}
\label{sec:BoundaryBagging}

While sparse regression on system \eqref{Eq:FullWFSystem} often yields accurate identification of PDE terms, in practice, the system tends to be dominated by its PDE component. 
This is due to the significantly greater number of test functions used in the PDE part compared to the ODE part, resulting in many more rows in system \eqref{Eq:LinSystemPDE} than in system \eqref{Eq:LinSystemODE}. 
As a consequence, the PDE residual becomes disproportionately weighted during optimization, causing the method to ``push'' errors into the boundary conditions. 
This typically leads to poor, and generally non-sparse, term selection performance in the boundary equations.

To address this, we introduce a cross-validation procedure for the ODE term selection.
This method is inspired by the library bagging technique used in Ensemble versions of SINDy and WSINDy \cite{FaselKutzBruntonEtAl2022ProcRSocA} (see comments in the discussion in \cref{Sec:Discussion}).
The difference here is in the method of selecting which terms to discard from the boundary process component of the library.
Specifically, we fix the learned source weights $\mathbf{w}^f$ and apply sparse regression to the modified system
\[
\mathbf{b}^{\text{ode}} - \Xi^f \mathbf{w}^f = \Xi^\beta \widehat{\mathbf{w}}^\beta,
\]
solving for $\widehat{\mathbf{w}}^\beta$.
We then compare the supports of the original and cross-validated boundary weights, $\text{supp}(\widehat{\mathbf{w}}^\beta)$ and $\text{supp}(\mathbf{w}^\beta) := \{i \in \{1,2,\dots,M^\beta\} : \textbf{w}^\beta_i \neq 0\}$. 
Often, the ODE-focused component returns a sparser subset of learned boundary terms, which tend to be more accurate representations of the true dynamics when tested against synthetic data. 
The terms that are not common between the methods are then removed from the boundary component of the library before refitting. 
We summarize the method in \cref{alg:WSINDyStructuredPop}.

\begin{algorithm}
\caption{WSINDyStructuredPop}
\label{alg:WSINDyStructuredPop}
\begin{algorithmic}[1] 
\Function{WSINDyStructuredPop}{$n_\text{data}$,$N_\text{data}$, $s$, $t$,$\{\phi_k\}$,$\{g_j,f_j,\beta_j\}$,$\dots$}
    \State Construct $G^g$, $G^f$, $\Xi^f$, $\Xi^\beta$, $\textbf{b}^\text{pde}$, and $\textbf{b}^\text{ode}$ from \cref{Eq:LinSystemPDE} and \cref{Eq:LinSystemODE}
    \State $\textbf{w}^g , \textbf{w}^f,\textbf{w}^\beta \gets $ MSTLS($[G^g,G^f,\Xi^f,\Xi^\beta]$,\textbf{b})
    \State $\widehat{\textbf{w}}^\beta \gets$ MSTLS($\Xi^\beta$, $\textbf{b}^{\text{ode}} - \Xi^f \textbf{w}^f$) 
    \If{supp$(\widehat{\textbf{w}}^\beta) \neq $ supp$({\textbf{w}}^\beta)$ }
    \If{supp$(\widehat{\textbf{w}}^\beta) \bigcap $ supp$({\textbf{w}}^\beta) \neq \emptyset$}
    \State idx $\gets$ supp$(\widehat{\textbf{w}}^\beta) \bigcap $ supp$({\textbf{w}}^\beta)$
    \Else
    \State idx $\gets$ supp$(\widehat{\textbf{w}}^\beta) \bigcup $ supp$({\textbf{w}}^\beta)$
    \EndIf
    \State $\textbf{w}^g , \textbf{w}^f,\textbf{w}^\beta \gets $ MSTLS($[G^g,G^f,\Xi^f,\Xi^\beta(:,\text{idx})]$,\textbf{b})
    \Else 
    \State $\textbf{w}^\beta \gets \arg\min_{\textbf{v} \in \{\textbf{w}^\beta,\widehat{\textbf{w}}^\beta\}} \textbf{R}([\textbf{w}^g;\textbf{w}^f;\textbf{v}])$
    \EndIf
    \State \Return $\textbf{w}^g , \textbf{w}^f,\textbf{w}^\beta$
    
\EndFunction
\end{algorithmic}
\end{algorithm}


\section{Results} \label{Sec:Results}
In the sections to follow, we will test the algorithm on several biologically motivated examples, which in particular focus on two well-studied models: age-structured population models \eqref{Eq:AgeStructured} and size-structured population models \eqref{Eq:SizeStructured}. 
We begin with a discussion on the noise assumptions and the performance metrics commonly used to assess these methods.

While standard explorations of the noise sensitivity of similar methods have traditionally involved using additive Gaussian noise, such an approach could result in the  artificial measurements of population data being negative, especially for large noise levels.
Therefore, to address this problem, we opt to use multiplicative log-normal noise to preserve the realistic nonnegative structure of the population data.
More precisely, we assume the elements in the dataset $\textbf{n}$ are of the form $n_j = \varepsilon_j n^\star_j$ such that $\varepsilon_j = \exp( z_j )$, where  $\{z_{j}\}_{j = 1}^{J}$ are i.i.d.  Gaussian variables with zero-mean and constant variance $\sigma^2$, $z_j \sim \mathcal{N}(0,\sigma^2)$.  
To measure the effective noise level from this distribution, we define the expected noise-to-signal ratio by 
\[\sigma_{NR} : = \frac{\mathbb{E}[(n-n^\star)^2]}{\|n^\star\|_{RMS}^2} = \mathbb{E}[(\varepsilon - 1)^2] = e^{\sigma^2}(e^{\sigma^2}-1) + (e^{\sigma^2/2}-1)^2.\]

One benefit of the WSINDy algorithm is that it can not only select proper model ingredients, but also provides an estimate of the linear coefficient present in front of the chosen functions.
This is a major benefit when population dynamics are concerned, as the coefficients in front of the total population variable, $N$, provide insight into critical parameters of the dynamics. 
For example, in the Verhulst population model, $\dot{N} = rN(1- N/k)$, $k$ is often interpreted as the carrying capacity of the environment. 
However, a large bias introduced by the log-normally distributed noise can cause the estimates of $N$ to be skewed and consequently  result in major errors in the estimates of these linear parameters, leading to large errors in the dynamics of the total population.
This is a critical observation in our setting as, due to the asymmetrical structure of the log-normal distribution, there is a bias encountered when calculating the total population from the noisy number density data. 
To correct for this, we approximate the variance of the population data using a local polynomial fit and dividing by the estimated expected value of the log-normal noise.\footnote{It is worth noting that this is unnecessary when the model ingredients do not depend on the total population, as \cref{Eq:FullWFSystem} would then only be affected by the dynamics of the total population and not the particular value of $N$.}
Precise details of this method are provided in the Appendix Section \ref{Sec:TotalPop}.

Table \ref{Tab:Metrics} summarizes the performance metrics used throughout the manuscript. 
To measure the accuracy of the recovered coefficients, we make use of two performance metrics commonly used in equation learning methods, denoted by $\textbf{E}_\infty$ and $\textbf{E}_2$.
Essentially, $\textbf{E}_\infty$ represents the upper bound of the relative error on the recovered  coefficients and $\textbf{E}_2$ provides information on the magnitude of the error on all coefficients.
Additionally, we will make use of the true positivity ratio denoted by \textbf{TPR}, where $TP$ represents the number of true positives, $FP$ the number of false positives, and $FN$ the number of false negatives identified by the method.

Finally, to measure the predictive power of the learned model, even when incorrect terms are learned, we define the training time interval by $[0,T_{\text{test}}]$ and whenever $T_{\text{test}} < T$, we measure the prediction accuracy of the model by simulating the evolution of the system using the learned model ingredients to construct a predicted solution $\tilde{n}$.
We then define the prediction error \textbf{E}$_p$ to be the relative $L^2$ norm of the true solution $n^\star$ and the predicted solution $\tilde{n}$ over the testing interval $(T_\text{test},T)$.
As we will see, when the library consists of similar functions, the learned equation could be analytically incorrect (i.e., $\textbf{TPR}< 1$); however, the learned model still fits the data as the difference between the model ingredients is small when weighted against the data.

\begin{table}[ht]
\centering
\begin{tabular}{c|c|c}
Label & Formula & Description \\ \hline
  $\textbf{E}_\infty$    & $\textbf{E}_\infty(\textbf{w}) := \max_{\{j : \textbf{w}^\star_j \neq 0\}} \frac{|\textbf{w}_j - \textbf{w}^\star_j|}{|\textbf{w}^\star_j|}$      &  Relative accuracy of true term estimators           \\ & & \\
  $\textbf{E}_2$    & $\textbf{E}_2 (\textbf{w}):=  \frac{\|\textbf{w} - \textbf{w}^\star\|_2}{\|\textbf{w}^\star\|_2}$ &  Magnitude of identified coefficients         \\ & & \\
   \textbf{TPR}   &     $\textbf{TPR}(\textbf{w}) := \frac{TP}{TP + FP +FN}$   &      True positivity ratio       \\ & & \\
   
 \textbf{E}$_p$     &   $\textbf{E}_p\textbf{(w)} := \frac{\|\tilde{n} - n^\star\|_{L^2((T_{\text{test}},T)\times \Omega)}}{\|n^\star\|_{L^2((T_{\text{test}},T)\times \Omega)}}$      &          Prediction error of learned model  \\ & & \\
 \textbf{R} & $\textbf{R(w)} := \frac{\|\textbf{b}-G\textbf{w}\|_2}{\|\textbf{b}\|_2}$ & Residual of the linear system  \\ 
 & &with learned weights  \\\hline
\end{tabular}
\caption{Performance metrics used throughout the paper.}
\label{Tab:Metrics}
\end{table}

\subsection{Construction of the library}
The construction of the matrix $G$ is the most computationally intensive and arguably the most important part of the algorithm detailed above. 
The question of which trial functions to include in the library is often best answered by expert insight into the population dynamics.
For example, depending on the population in question, age-structured mortality functions have been modeled using exponential and polynomial functions \cite{AvraamArnoldVasievaEtAl2016Aging}. 
Therefore, if possible, one should treat the libraries as an array of distinct hypothesis functions the WSINDy algorithm will then select between. 
It is worth pointing out that while WSINDy can estimate linear parameters, all nonlinear parameters must currently be given in the library.
Extending weak-form algorithms such as WSINDy and WENDy to estimate these nonlinear parameters is still an active research area with some sparse recent advancements \cite{DucciKouyateReuterEtAl2025JChemPhysa,RummelMessengerBeckerEtAl2025arXiv250208881}, and so we postpone this update to our method for future work.  

We will construct our libraries as if we are well informed regarding the general shape of the true model ingredients, e.g., we will assume that the fecundity rate is approximately Gaussian.
Accordingly, we will make use of functions commonly found in the literature of structured populations including polynomial $f_\text{poly}(s;p) = s^p$, exponential $f_\text{e}(s;k) = e^{ks}$, sigmoidal $f_\text{sig}(s;s_c,k) = \frac{1}{1+\exp(-k(s-s_c))}$, and Gaussian $f_\text{Gauss}(s;\mu,\sigma) = \exp({-(s-\mu)^2/(2\sigma^2)}).$ 
Additionally, we assume the dependence on the total population is multiplicative, i.e. $f(s,N) = f_1(x)f_2(N)$, as is common in the literature \cite{AcklehLyonsSaintier2020MathBiosciEng,GurtinMaccamy1974ArchRationalMechAnal}, however, we remark this is not necessary in general.

\subsection{Linear Example Problems}
We begin by assessing the method's ability to identify linear population models, that is, where the dynamics are independent of the total population size.
We take four representative linear structured population models that differ in their transport, source, and boundary terms. 
The specific forms of the true functions used in each example are summarized in Table \ref{Table:Example problemsAge}. 
In these examples, the true transport (growth/aging) term $g^\star(s)$, source (death) term $f^\star(s,n)$, and boundary (birth) term $\beta^\star(s)$ are chosen to be simple, but commonly used, functions of the structuring variable $s$ (e.g., age or size). 
The specific libraries used in each example are listed in Table \ref{Table:Ex_Libraries}.

\begin{table}[ht]
\centering
\begin{tabular}{ccccc}
\textbf{Example} & \textbf{$g^\star(s,N)$} & \textbf{$f^\star(s,n,N)$} & \textbf{$\beta^\star(s,N)$}                 & \textbf{$(0,T)\times\Omega$} \\ \hline
 L.1            & 1                       & $-0.3n$                  & $0.4$           & $(0,10)\times[0,15]$            \\ 
 L.2            & $1 $                      & $-0.1\exp(0.08s)n$ & $f_{\text{Gauss}}(s;10,5)$ & $(0,10)\times[0,15]$             \\ 
 L.3 & $0.2(3-s)$ & $-0.3 s\,n$ & $f_{\text{sig}}(s;1,2)$ & $(0,10)\times[0,3]$ \\ 
 L.4 & $s(1-0.25s)$ & $-0.7\exp(0.2s)n$ &$0.6s$ &$(0,10) \times [1,4]$\\ \hline
\end{tabular}
\caption{\label{Table:Example problemsAge}The critical parameters for the linear example problems.}
\end{table}

\begin{table}[ht]
    \centering
    \begin{tabular}{cccc}
    \textbf{Example} & \textbf{Transport Terms} & \textbf{Source Terms} & \textbf{Boundary Terms}               \\ \hline
         L.1 & $\{s^p\}_{p = 0}^2$ & $\{s^p n\}_{p = 0}^3$, &  $\{s^p\}_{p = 0}^3 $, \\
         L.2 & $\{s^p\}_{p = 0}^2$ & $\{\exp(0.08(4k+1) s)n \}_{k=0}^4$ & $ \{f_\text{Gauss}(s;5k,5)\}_{k = \{1,2,3\}}$\\
         L.3 & $\{s^p \}_{p = 0}^2$ & $\{s^p n\}_{p = 0}^4$ & $ \{f_\text{sig}(s;k,2)\}_{k = 1}^4$ \\ 
         L.4 & $\{s^p \}_{p = 0}^2$ & $\{\exp((0.2+0.5k) s)n \}_{k=0}^3$ &  $\{s^p\}_{p=1}^4$\\
        \hline
    \end{tabular}
    \caption{Libraries for the linear example problems in Table \ref{Table:Example problemsAge}.}
    \label{Table:Ex_Libraries}
\end{table}

\subsubsection{The effect of noise}\label{Sec:Noise}
\paragraph{No Noise:}
In Table \ref{Table:NoNoiseMetrics}, we present the coefficient errors and residual from the learned coefficients in the ideal case $\sigma_{NR} = 0$. This provides a baseline test of the algorithm where all of the errors are numerical in nature and not due to added noise. 

\begin{table}[ht]
    \centering
    \begin{tabular}{c|c|c|c|c}
    \textbf{Label} & $\textbf{E}_\infty$ & $\textbf{E}_2$ & $\textbf{R}$ &$\textbf{E}_p$              \\ \hline
        Ex L.1 & $9.6\mathrm{e}{-06}$ & $6.1\mathrm{e}{-06}$ &  $4.1\mathrm{e}{-06}$ &$5.3\mathrm{e}{-04}$ \\
        Ex L.2 & $5.6\mathrm{e}{-04}$ & $3.9\mathrm{e}{-04}$ & $ 1.1\mathrm{e}{-05}$
        & $ 3.4\mathrm{e}{-03}$\\
        Ex L.3 & $2.9\mathrm{e}{-04}$ & $2.6\mathrm{e}{-04}$ & $ 5.2\mathrm{e}{-05}$& $ 3.9\mathrm{e}{-03}$\\
        Ex L.4 & $8.2\mathrm{e}{-04}$ & $5.2\mathrm{e}{-04}$ & $2.3\mathrm{e}{-04}$ &$5.7\mathrm{e}{-04}$\\ \hline
    \end{tabular}
    \caption{$\textbf{E}_\infty$, $\textbf{E}_2$, residual, and prediction error for the test cases and libraries presented above with $T_{\text{test}} = 0.5T$ and $\sigma_{NR} = 0$.
    All examples for the no noise case have $\textbf{TPR}=1$.}
    \label{Table:NoNoiseMetrics}
\end{table}

\begin{figure}[ht]
    \centering
    \includegraphics[width=1\linewidth]{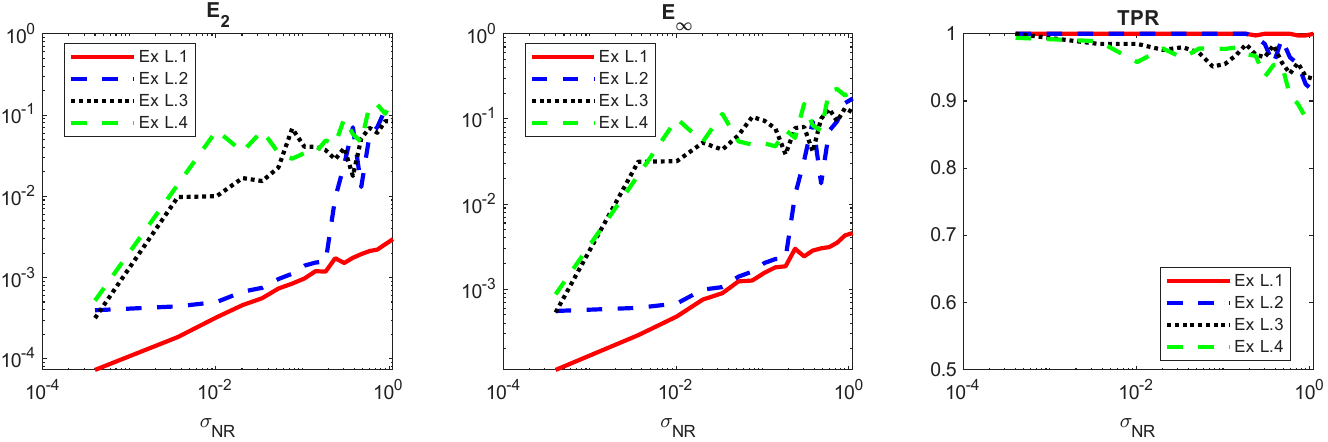}
    \includegraphics[width=1\linewidth]{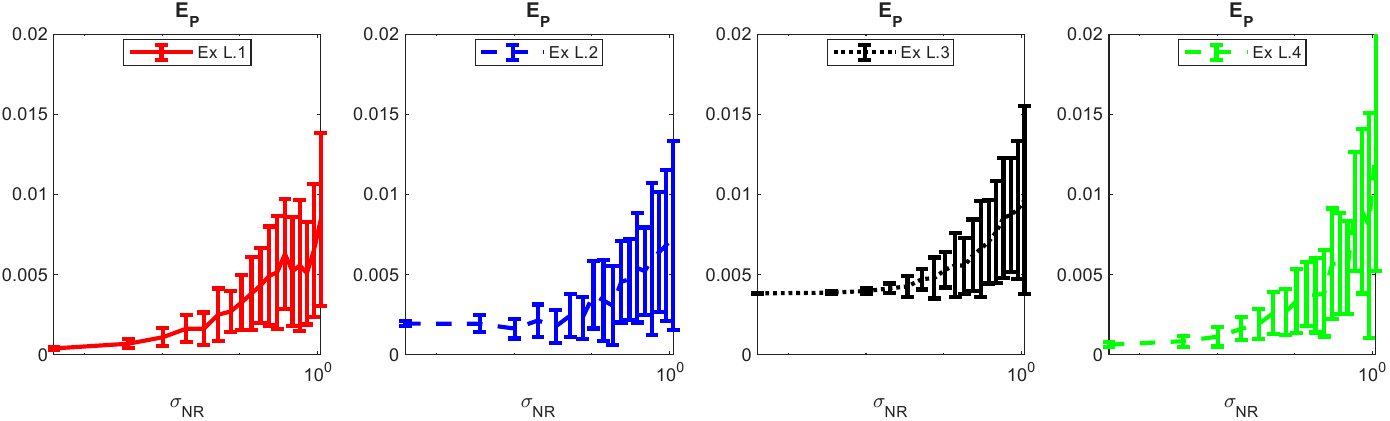}

    \caption{\textbf{Top:} Average performance metrics for the linear models in Table \ref{Table:Example problemsAge} using the libraries in Table \ref{Table:Ex_Libraries} plotted against various noise levels using 100 dataset realizations at each noise level.\\
    \textbf{Bottom:} Average and inner quartile range of the prediction error, $\textbf{E}_p$,  for the learned coefficients.}
    \label{fig:LinearPerformaceMetrics}
\end{figure}

\begin{figure}[ht]
    \centering    
    \includegraphics[width = 1\linewidth,height = 0.3\linewidth]{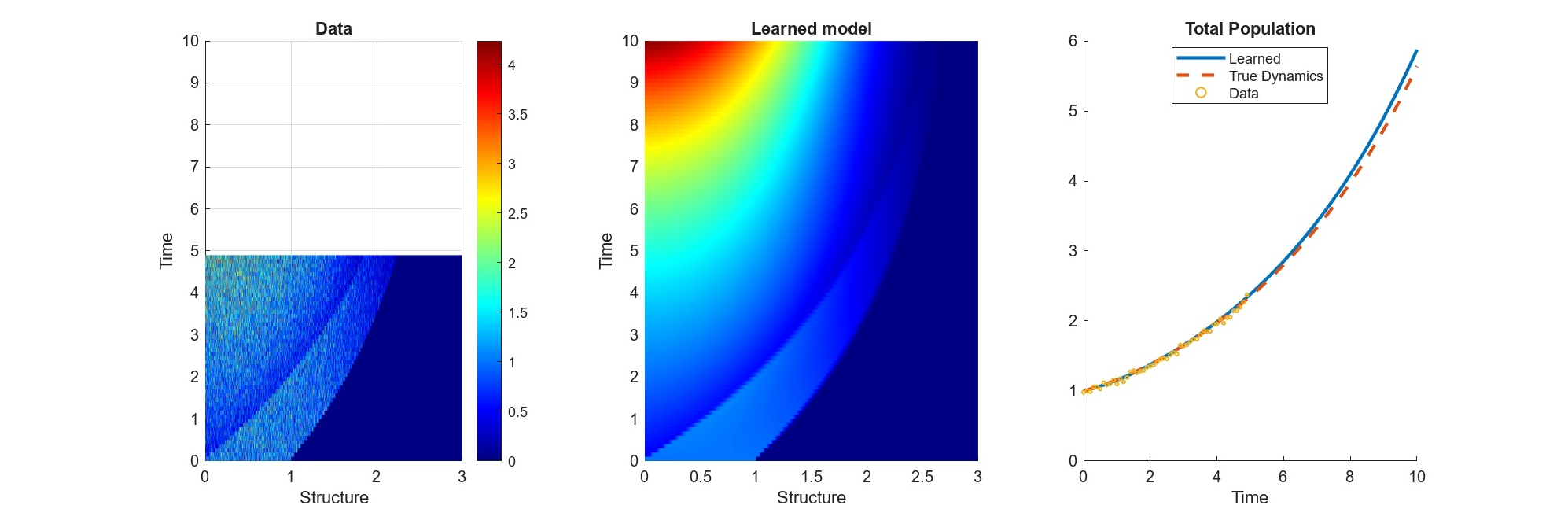}
    \caption{Typical results of using the WSINDy algorithm for example problem L.3 in Table \ref{Table:Example problemsAge} with the library presented in Table \ref{Table:Ex_Libraries}, $\sigma_{NR} = 0.66$, and 50 points in time. The prediction error and \textbf{TPR} of the learned model approximately 0.035 and 0.8, respectfully.}
    \label{fig:TypicalL3}
\end{figure}

\paragraph{Noise:} 
In Figure \ref{fig:LinearPerformaceMetrics}, we present the measured performance metrics for the linear models presented in Table \ref{Table:Example problemsAge} using 3000 points in the structural variable and 500 points in time.
It is worth noting that as noise is added to the data, the \textbf{TPR} of the learned coefficients expectantly drops off. 
However, the prediction error maintains reasonable levels proportional to the noise level, indicating that the method can consistently learn effective models using the library terms even if they are analytically incorrect.  
This is demonstrated for example L.2 in Figure \ref{fig:TypicalL3} where the learned model has a \textbf{TPR} value of 0.8, but the model fits the data well and maintains a reasonable prediction level.
We provide similar examples of this with the other models in Section \ref{Sec:AppendPlotsandFigs} of the Appendix.
Additionally, the average run time of each realization and model recovery procedure was less than 10 seconds and can be easily run on any modern laptop.

\subsubsection{Distinguishability of Library Terms}\label{SubSec:Distinguish}

In this section, we computationally investigate the distinguishability of the candidate functions within the library.\footnote{How to mathematically characterize and evaluate the distinguishability of a given library for a given data set using sparse regression-based equation learning is currently an open question (see the discussion in \cref{Sec:Discussion}).} 
Specifically, we examine how effectively the algorithm can distinguish between similar functions of the structural variable during the recovery process. 
This question is central to the reliability of sparse identification methods such as WSINDy, especially when multiple candidate terms produce similar features with respect to the data.

To explore this, we focus on Example L.2 and, to isolate the effects of term distinguishability, we simplify the transport component by restricting the library to include only the true transport term, $g^\star(s) = 1$. 
This ensures that the algorithm is not influenced by potential ambiguity in the transport dynamics, allowing us to study the source and boundary components in isolation.

We consider two test cases that assess the distinguishability of the source and boundary terms, respectively:
\paragraph{Case 1 (Source term):}
We fix the boundary term to include only the true birth rate, $f_{\text{Gauss}}(s;10,5)$. 
For the source term, we construct a series of libraries containing perturbed exponential functions of the form:
\[
\left\{ \exp\left((0.08 + \delta i)s\right)n \right\}_{i = -k}^k,
\]
where $\delta$ controls the spacing between candidate terms and $k$ determines the total number of alternatives. 
This setup allows us to assess how sensitive the algorithm is to small variations in source dynamics and how well it can single out the correct term among closely spaced candidates.
Additionally, the structure of the libraries guarantees that the true function $f^\star(s,n) =  -0.1\exp(0.08s)n$ is always present for every choice of $\delta$ and $k$.

\paragraph{Case 2 (Boundary term):}
We now fix the source term to include only the true function, $f^\star(s,n) = -0.1 \exp(0.08s)n$, and assess distinguishability in the boundary term. 
To this end, we consider libraries of Gaussian profiles of the form:
\[
\left\{ f_{\text{Gauss}}(s; 10 + \delta i, 5) \right\}_{i = -k}^k,
\]
where the means of these functions vary over different spacing.

\begin{figure}[ht]
    \centering
    \includegraphics[width=0.75\linewidth,height=0.3\linewidth]{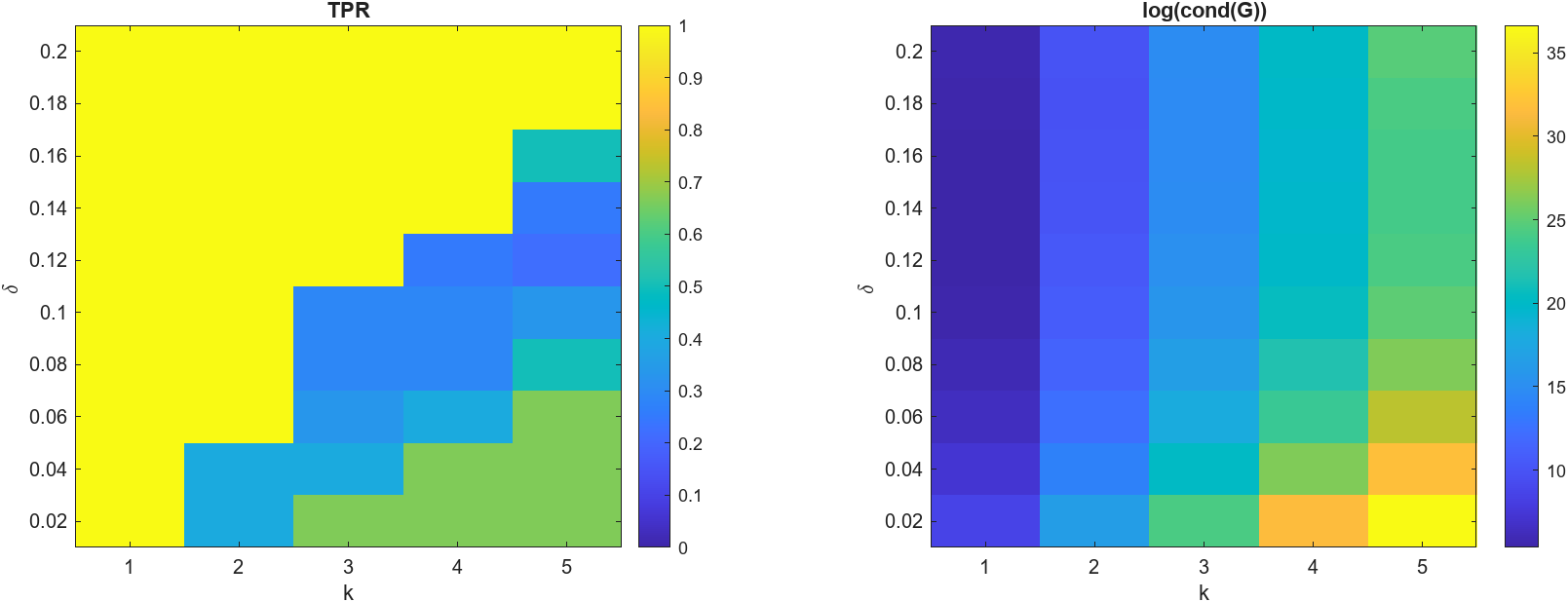}
    \includegraphics[width=0.75\linewidth,height=0.3\linewidth]{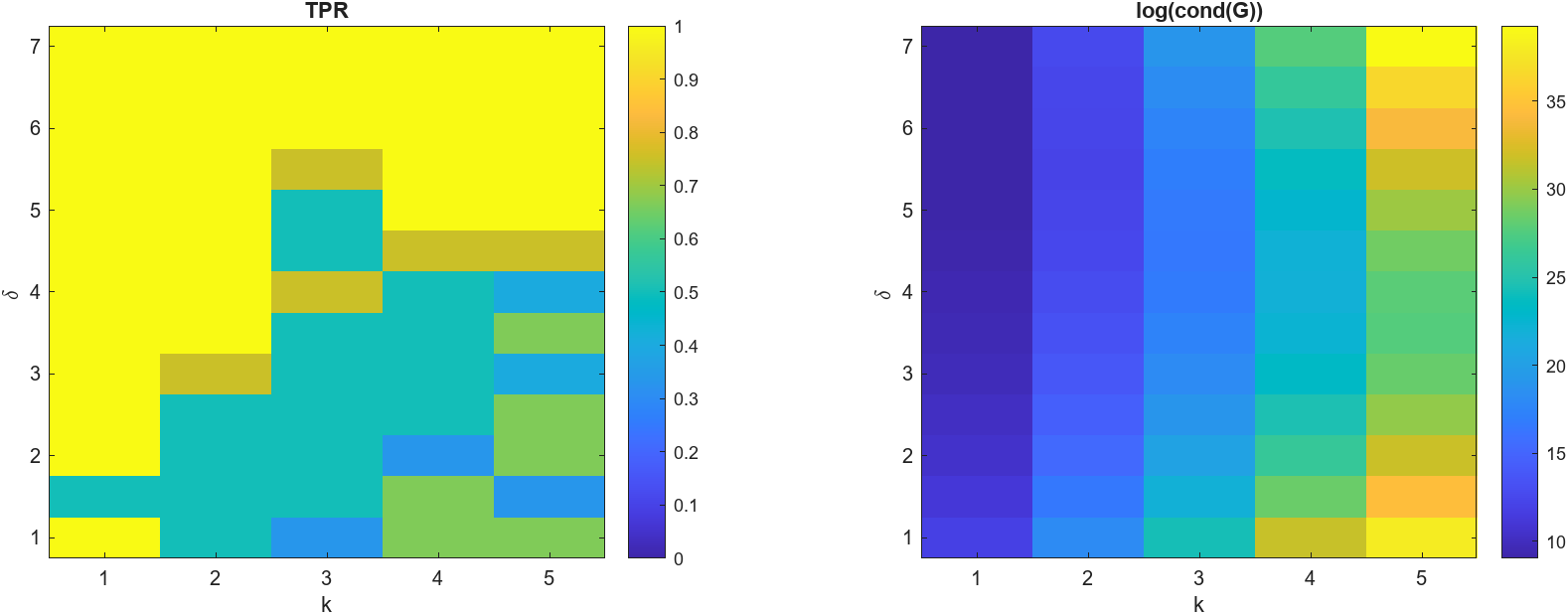}
    \caption{TPR and condition number over different choices of library parameters for case 1 (top) and case 2 (bottom).  }
    \label{fig:libTPR}
\end{figure}

We present the true positive ratio (\textbf{TPR}) results for both cases in Figure~\ref{fig:libTPR}, assuming no noise ($\sigma_{NR} = 0$). 
The corresponding condition numbers of the libraries are also shown to indicate how difficult it is to distinguish terms based on their linear dependence.
From these results, we observe a general trend: as the spacing $\delta$ decreases and candidate terms become increasingly similar, the algorithm's ability to correctly identify the true term deteriorates. 
In particular, the \textbf{TPR} drops sharply once the candidate functions are sufficiently close in structure, indicating that one should be aware and cautious of the similarity between the trial functions.

\subsection{Nonlinear Example Problems}
\begin{figure}
    \centering
    \includegraphics[width=1\linewidth]{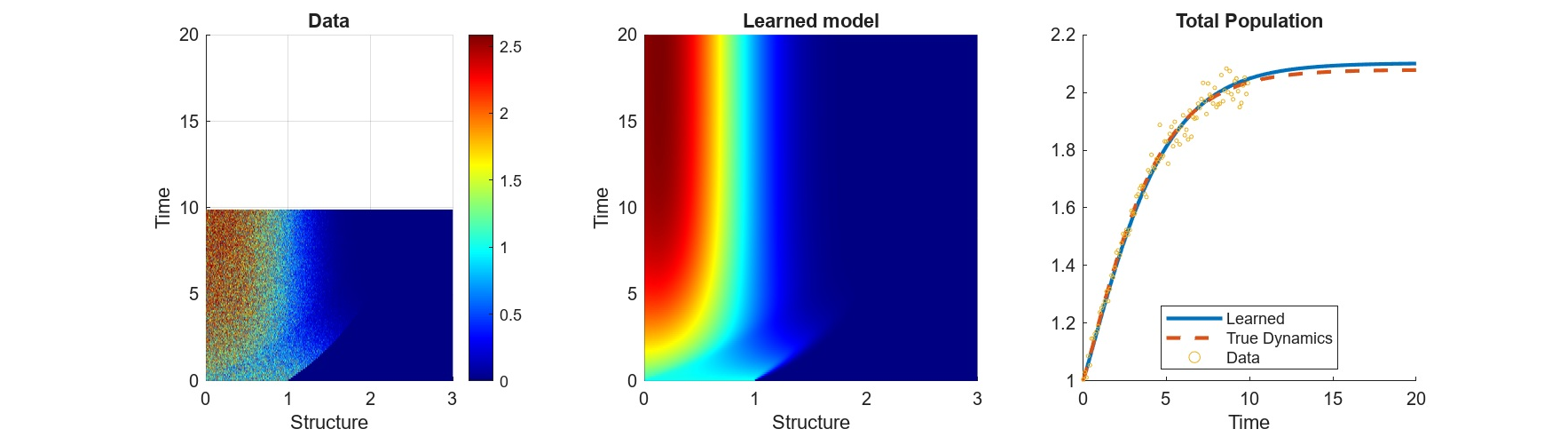}
    \caption{A typical run of Example NL.2 with $\sigma_{NR} =0.66$ and 50 points in time. The prediction error and \textbf{TPR} of this learned model are 0.004 and 1, respectively. }
    \label{fig:NonlinearExample}
\end{figure}

In this section, we consider the more complex case of nonlinear models, where the governing biological processes depend on the total population, $N$. 
We will consider the following two examples in this context:
\begin{table}[h]
\centering
\begin{tabular}{ccccc}
\textbf{Example} & \textbf{$g^\star(s,N)$} & \textbf{$f^\star(s,n,N)$} & \textbf{$\beta^\star(s,N)$}                 &  \\ \hline
NL.1            & 1                       & $- 0.6N\,n$                  & $1.5\exp(-0.25N)$                 \\ 
 NL.2            & $0.2(3-s)\exp(-0.25N) $ & $-0.3s(1+N)n$ & $2f_{\text{sig}}(s;1,2)\, (1-0.1N)$            \\ \hline
\end{tabular}
\caption{\label{Tab:ExamplesNonLin}The critical parameters for the nonlinear example problems. The domains for both problems are given by $(0,20)\times[0,15]$  and  $(0,20)\times[0,3]$, respectively. }
\end{table}
We use libraries similar to those in Table \ref{Table:Ex_Libraries} with the addition of the variable $N$ and so we omit the libraries for brevity. 
We provide an example of using the WSINDy method for recovering Ex NL.2 in Figure \ref{fig:NonlinearExample}.

While the method certainly can recover the true dynamics, the introduction of the dependence on the total population comes with several challenges for model discovery using WSINDy which we would like to discuss.
First, nonlinear models often converge relatively quickly to an equilibrium, especially when the dynamics occur over long time scales. 
As a result, the data becomes nearly stationary, making it difficult to identify the underlying dynamical system.
Second, convergence to equilibrium leads to many functions of the total population $N$ appearing structurally similar, which complicates their identification within the candidate function library. 
This indistinguishability is further exacerbated by WSINDy’s flexibility in adjusting the linear coefficients of selected terms. 
In particular, the algorithm may favor effective linear approximations at the cost of omitting the correct nonlinear structure.

\begin{figure}[h]
    \centering
    \includegraphics[width=0.45\linewidth]{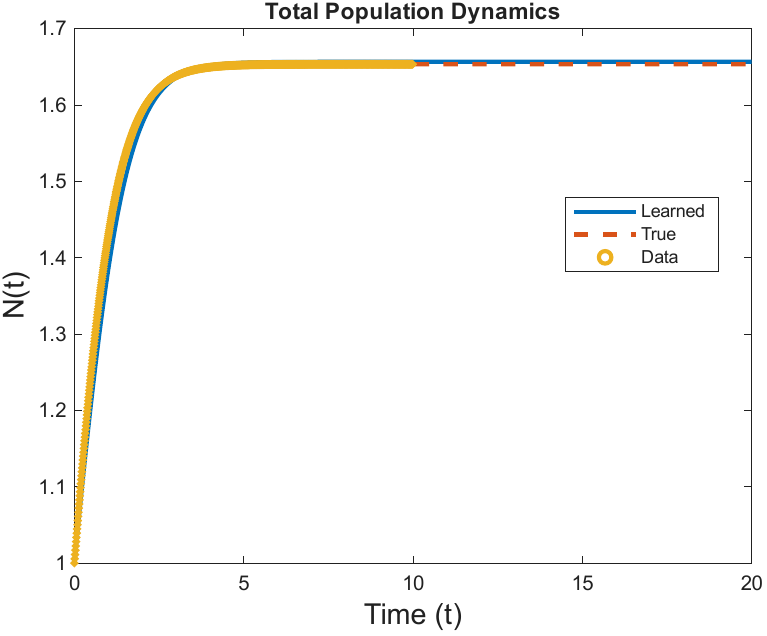}
    \includegraphics[width=0.45\linewidth]{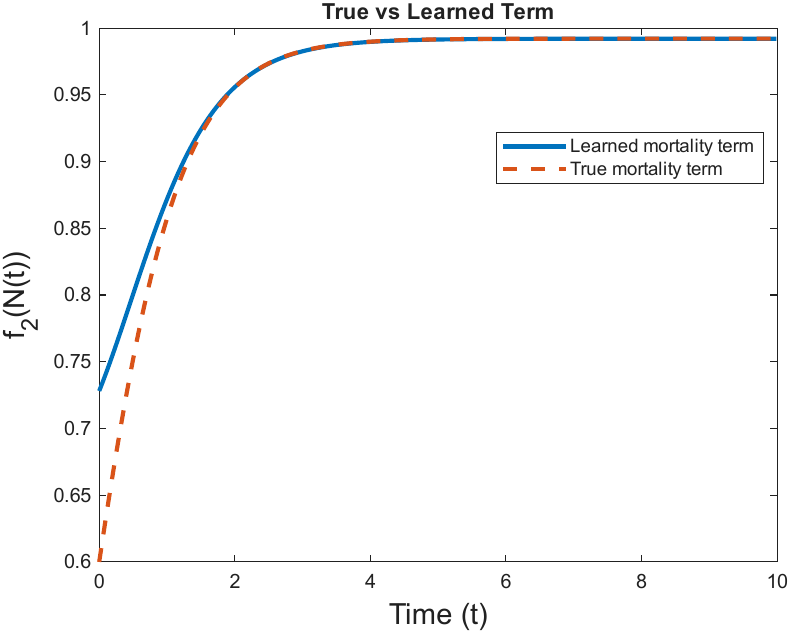}
    \caption{\textbf{Left:} The learned and true dynamics of the total population, $N$, plotted with the training data with noise ratio $\sigma_{NR} \approx 0.02$.
    \textbf{Right:} The learned nonlinear structure (of the form $a + bN^3$) plotted with the true structure over the training time interval.}
    \label{fig:NL1TrueVsLearned}
\end{figure}

To demonstrate these effects, we focus first on example problem NL.1 where the natural approach to recovering the true source term would be to use a library of monomials such as $\{N^kn\}_{k=0}^3$, however, even at very low levels of noise, the monomial library has distinguishability issues. 
This is demonstrated in Figure \ref{fig:NL1TrueVsLearned} where the method learns an effective, but incorrect, mortality term.
The algorithm struggles to differentiate between candidate terms because the key differences among them occur near $t = 0$, where the compactly supported test functions are near zero.\footnote{At low noise levels, this issue can often be mitigated by reducing the radius of the test function support. However, doing so typically degrades performance at moderate to high noise levels.}

\begin{figure}[h]
    \centering
    \includegraphics[width=\linewidth]{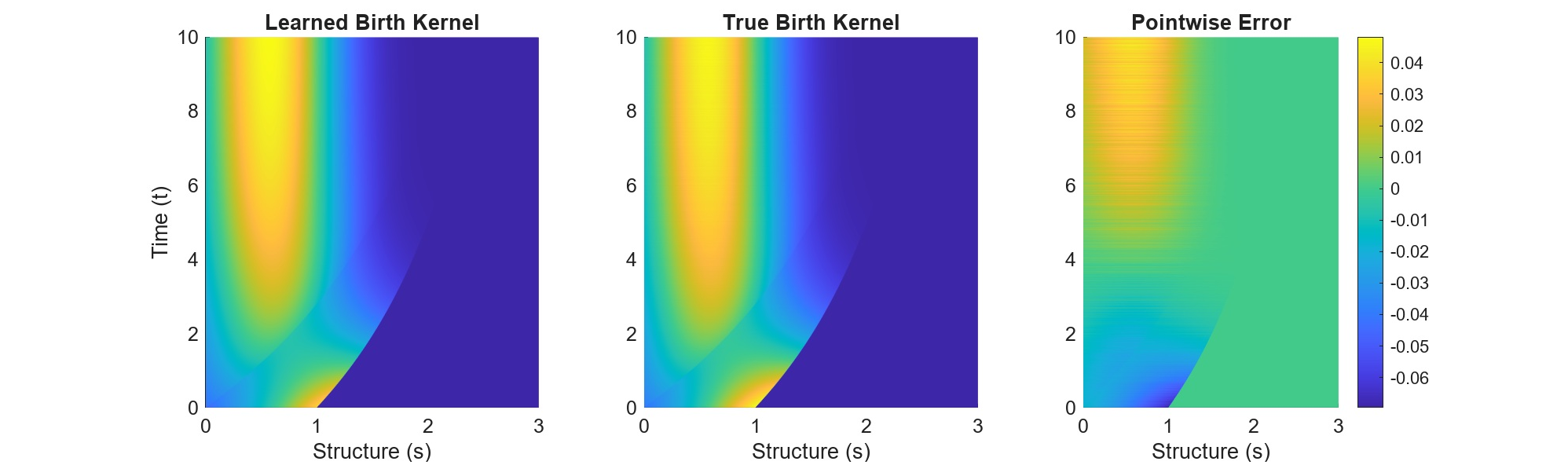}
    \caption{From left to right, we present the learned linear birth kernel $\beta(s) n^\star(t,s)$, the true nonlinear birth kernel $\beta^\star(s,N(t)) n^\star(t,s)$, and their pointwise error over the training time.}
    \label{fig:NL2LearnedVsTrueBirth}
\end{figure}

Next, we consider example NL.2, which incorporates a commonly used logistic-type nonlinearity in the birth rate, $\beta^\star$. 
When the data is significantly corrupted by noise (e.g., noise level $\sigma_{NR} > 0.20$), the algorithm often replaces the nonlinear structure with an effective linear approximation. 
To illustrate this behavior, Figure \ref{fig:NL2LearnedVsTrueBirth} compares the learned and true birth rates, both multiplied by the true population density.
In this case, the selected birth rate was given by $\beta(s) = 1.6638 f_{\text{sig}}(s;1,2) \approx 2(1-0.1\bar{N}) f_{\text{sig}}(s;1,2)$ where $\bar{N}$ is the average value of $N$ over the training time.
This structure, along with the corresponding plot, demonstrates that although the learned linear birth kernel is analytically incorrect, it is still  effective (and sparser) in capturing the birth process from the data set.

\subsection{Application to Real-World Age-Structured Data}\label{Subsec:RealWorldData}

\begin{figure}
    \centering
    \includegraphics[width=1\linewidth,height = 0.35\linewidth]{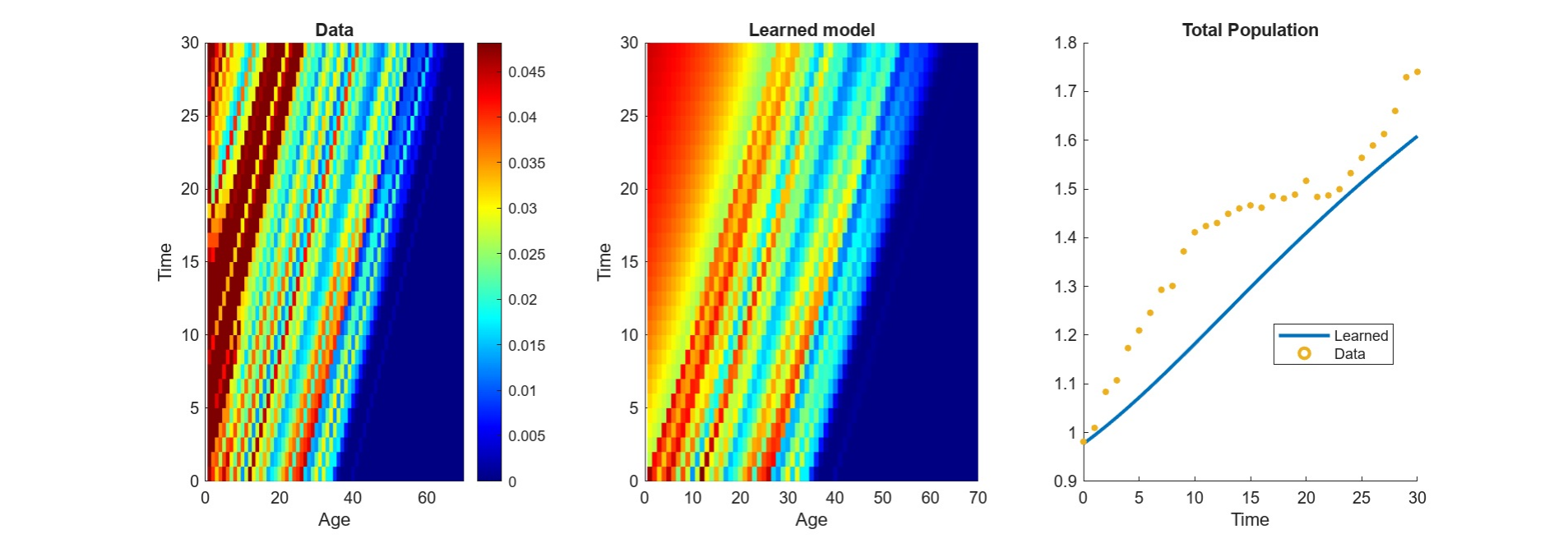}
    \includegraphics[width=1\linewidth]{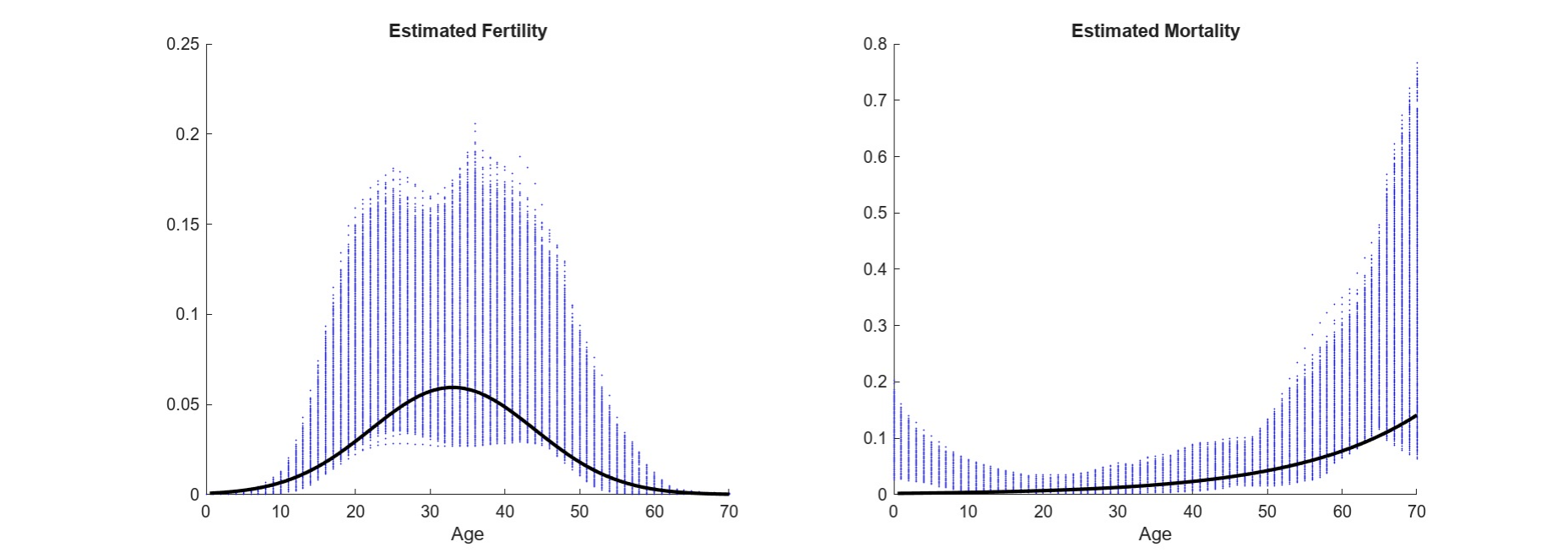}
    \caption{\textbf{Top:} WSINDy-reconstructed age-structured population dynamics of the semi-captive Asian elephant population over the 31-year period. 
    \textbf{Bottom:} Estimated age-specific fertility and survival functions inferred by WSINDy (solid lines), shown alongside those derived from classical matrix projection modeling in \cite{JacksonMarHtutEtAl2020JournalofAnimalEcology} (blue dots).}
    \label{fig:RealData}
\end{figure}

In this section, we demonstrate the utility of WSINDy on a real-world, age-structured population data set that documents the demographic dynamics of a semi-captive population of Asian elephants (\textit{Elephas maximus}) using a data set publicly available via the Dryad digital repository \cite{JacksonMarHtutEtAl2020} and analyzed from a discrete time modeling perspective in \cite{JacksonMarHtutEtAl2020JournalofAnimalEcology}.
The dataset records the age and demographic status of individual elephants, enabling us to construct an empirical approximation of the age-structured population density. 
To achieve this, we aggregate individuals into discrete age classes by binning at a resolution of one year, yielding a structured data array with 81 age classes and 31 temporal snapshots.
Although the temporal resolution is relatively coarse for data-driven model discovery, WSINDy is still capable of identifying interpretable and biologically plausible model terms. 
The resulting equation was of the form
\begin{subequations}\label{Eq:Elephant equation}
   \begin{empheq}[left=\empheqlbrace]{align}
       &\p_t n(t,a) + \alpha \p_a n(t,a) = -c\exp(0.06(a-85))n(t,a),\notag\\
       &n(t,0) = \int_0^\infty b f_{\text{Gauss}}(a;33,11) n(t,a) \diff{a},\notag
   \end{empheq} 
\end{subequations}
with $(\alpha,b,c) \approx (0.9184,0.3468,0.0595)$.
Figure \ref{fig:RealData} presents the results of applying WSINDy to the elephant data. 
The top panel compares the reconstructed population dynamics to the binned representation of the data, both at a density level and at the level of the total population.
The bottom panel compares the inferred survival and fertility functions to those derived from the matrix population modeling as reported in \cite{JacksonMarHtutEtAl2020JournalofAnimalEcology}.
These results show that the WSINDy-inferred model ingredients fall within reasonable ranges when compared to traditional modeling techniques.

\section{Discussion}\label{Sec:Discussion}
The framework of Scientific Machine Learning (SciML) aims to merge methods from scientific computing with those from machine learning to generate accurate and interpretable data-driven models. In this work, we have further developed Weak form SciML (WSciML) to learn PDE-based structured population models. To the best of the authors' knowledge, this effort is the first to use SciML methods to learn this class of population models. 
To this end, we have presented the first iteration of a weak-form equation learning algorithm for structured population equations and have demonstrated the potential of the method on both synthetic and real datasets.

As with any approach, WSciML methods have advantages and disadvantages.  In this section, we discuss both the advantages as well as the current limitations.  
The most notable advantage is the capability of the method  to directly learn the governing equation, bypassing the time-consuming iteration between model form creation, numerical approximation, and validation after  parameter fitting.  
It is of course, still necessary to carefully curate a library of mechanistically justified model features from which to build the model. Given a suitable library, the WSciML process allows researchers to simultaneously test multiple interpretable models for, e.g., structuring relationships, instead of investigating one hypothesis at a time.
It is of course, still necessary to carefully curate a library of mechanistically justified model features from which to build the model. Given a suitable library, the WSciML process allows researchers to simultaneously test multiple interpretable models for, e.g., structuring relationships, instead of investigating one hypothesis at a time. As a result, WSciML methods can be orders of magnitude faster than the traditional model selection and parameter inference methods \cite{BortzMessengerDukic2023BullMathBiol}.

Another important advantage of WSciML methods is their robustness to noise.  
As demonstrated in Figure \ref{fig:LinearPerformaceMetrics} (as well as in other previous publications \cite{MessengerBortz2021JComputPhys,MessengerBortz2021MultiscaleModelSimul,BortzMessengerDukic2023BullMathBiol, BortzMessengerTran2024NumericalAnalysisMeetsMachineLearning,MessengerBortz2024IMAJNumerAnal}), the weak form integral transform offers a SciML method which is highly robust to noise. 
Furthermore, while there is evidence that for certain classes of models, WSciML methods work very well with sparse data \cite{MessengerBortz2024IMAJNumerAnal}, structured population models present their own challenges. 
In particular, this (initial) version of using WSINDy to learn structured populations is moderately sensitive to the number of structure classes in the binned data.  
It is also critical that the time series of the structured data include observations that are sufficiently rich in information content so as to allow statistically warranted inclusion or pruning of terms in the library. However, a precise quantification of the needed \emph{richness} is beyond the scope of this work and will be the focus of future efforts.
Regardless, with sufficient data, it's clear that WSINDy performs well at selecting effective models that not only fit the data but also possess solid predictive capabilities. 

Several studies have also examined how data resolution affects the distinguishability of library terms. 
A common approach to addressing this issue involves using multiple datasets with varying initial conditions to gain a more comprehensive understanding of the system's dynamics \cite{VaseyMessengerBortzEtAl2025JournalofComputationalPhysics,LyuGalvanin2024ComputerAidedChemicalEngineering}. 
In our setting, data resolution appears to be closely linked to the support of the population density: a larger support over the time series (relative to the size of $\Omega$) tends to yield improved distinguishability across the library terms.
A more rigorous study of this effect is left as future work.

As mentioned in \cref{sec:BoundaryBagging}, for our structured population models, a naive sparse regression of the combined ODE/PDE model in \eqref{Eq:FullWFSystem} resulted in correctly learning the PDE at the expense of incorrectly learning the ODE.  For this class of problems, it would thus be natural to consider the weak form version of the Ensemble-SINDy (E-SINDy) approach by Fasel et al., \cite{FaselKutzBruntonEtAl2022ProcRSocA}. However, we found that E-WSINDy needed a (relatively) larger library to be effective, which resulted in ill-conditioned matrices (for our example problems in this paper). Accordingly, in our example problems, we made use of relatively small libraries, which necessitated a rather restrictive set of hyperparameters in the traditional E-WSINDy method.
Because of this, we opted not to make use of the original E-WSINDy method and used our own extension, based on cross-validation (see \cref{sec:BoundaryBagging}).
In the case of a much larger library, however, E-WSINDy would remain an efficient method for  model discovery.

Lastly, we explored potential improvements to the method by focusing on two aspects.
First, in this introductory study, we construct the library using a collection of structurally identical trial functions, each parameterized by different (but often similar) values. 
As demonstrated in Section \ref{SubSec:Distinguish}, including a large number of such near-redundant functions can significantly increase the condition number of the weak form matrix $G$.
A high condition number amplifies numerical instability and can impair the algorithm's ability to reliably distinguish between candidate terms. 
This, in turn, may lead to poor model selection, where incorrect terms are favored due to small differences in fit quality being exaggerated by ill-conditioning.
A possible avenue to remedy this issue is to iterate the selection step defined in Section \ref{Sec:WSINDy} with a library adjustment step where the parameters in the selected trial functions are modified in such a way as to better fit the data \cite{DucciKouyateReuterEtAl2025JChemPhysa}.
Additionally, in the work above, the regression steps do not take into account any natural intuition from the model.
Indeed, depending on the application at hand, the method could be tailored to the dynamics by including natural modeling constraints into the regression.
Different applications of such models have a variety of natural restrictions on the model ingredients, such as a decaying survival probability in animal populations, symmetric division in cell populations, or mass conservation in structured coagulation-fragmentation equations. 
Adding these constraints to the regression can significantly improve the interpretability of the models and may result in more accurate identification of the dynamics at higher noise levels or lower data resolutions.

Second, in the case where the structure of the true model ingredients is completely unknown, one may seek to make use of nonparametric representations of the trial functions and aim to ``build up" the true ingredients in a similar manner to forward-solver least-squares techniques \cite{BanksBotsfordKappelEtAl1991JMathBiol,BanksSuttonClaytonThompsonEtAl2011BullMathBiol}.
One can construct such a library using nonparametric functions in the same way as presented in Section \ref{Sec:WSINDy} and use non-sparse regression techniques to find effective models.
This method, however, allows the algorithm much freedom in selecting model ingredients and, as such, can result in highly variable and even unrealistic representations of the model ingredients and would possibly be better handled by parameter estimation methods such as WENDy.
This is an active frontier of research.

\paragraph{Data Availability:}
All code used in this manuscript will be available on the repository \\ \href{https://github.com/MathBioCU/WSINDyStructuredPopulations}{github.com/MathBioCU/WSINDyStructuredPopulations}.

\paragraph{Acknowledgements:}
Research reported in this publication was supported in part by the NIGMS Division of Biophysics, Biomedical Technology and Computational Biosciences (grant R35GM149335); NSF Division of Molecular and Cellular Biosciences MODULUS (grant 2054085); NSF Division Of Environmental Biology (EEID grant DEB-2109774), and NIFA Biological Sciences (grant 2019-67014-29919). 
This work also utilized the Blanca condo computing resource at the University of Colorado Boulder. Blanca is jointly funded by computing users and the University of Colorado Boulder.
The authors would also like to thank D.~Messenger (Los Alamos National Labs) and N.~Heitzman-Breen (CU Boulder) for discussion regarding the implementation of the method.

\nolinenumbers

%
%
%
\bibliographystyle{ieeetr}
\bibliography{StructuredPopulationModels.bib}

\appendix
\section{Simulation details}
Due to the histogram-style interpretation of the data, it is natural to use a finite-volume method to construct artificial data for the example problems.
To this end, the artificial examples presented in Table \ref{Table:Example problemsAge} are simulated using a minmod flux-limiter discretization which has been previously used and analyzed for a variety of size-structured models \cite{AcklehChellamuthuIto2015MathBiosciEng,AcklehLyonsSaintier2020MathBiosciEng,AcklehLyonsSaintier2023MBE,AcklehMa2013NumerFunctAnalOptim} and we refer the reader to these works for more details and convergence properties of the scheme.
For simplicity, we present the method here for the size-structured population model \eqref{Eq:SizeStructured} and remark that the age-structured case follows with $g \equiv \alpha$.
Let $\Delta x >0$ be a given mesh size and let $\Lambda_i := (x_i - \frac{1}{2}\Delta x, x_i + \frac{1}{2}\Delta x]$ represent uniformly placed cells with midpoints $x_i = (i-\frac{1}{2})\Delta x$ with $i = 1,2,\dots,I$.
We then discretize the initial condition as $n^0_i := \frac{1}{|\Lambda_i|}\int_{\Lambda_i} n(0,x) \diff{x}$ and, through integrating \eqref{Eq:SizeStructured} over the domains $\Lambda_i$, arrive at the following system of differential equations which describe the evolution of the volumes $n_i(t) = \frac{1}{|\Lambda_i|}\int_{\Lambda_i} n(t,x) \diff{x}$,
\begin{subequations}\label{Eq:FV_scheme}
\begin{empheq}[left=\empheqlbrace]{align}
\frac{\diff{}}{\diff{t}}n_i &= -\frac{1}{\Delta x} [\mathcal{F}_{i+1/2}-\mathcal{F}_{i-1/2}] - d[\vec{n}]_i n_i, \\
g[\vec{n}]_0 n_0 &= \Delta x \left(\frac{3}{2}\beta[\vec{n}]_1 n_1 + \sum_{i=2}^{I-1}\beta[\vec{n}]_i n_i + \frac{1}{2}\beta[\vec{n}]_I n_I \right),
\end{empheq}    
\end{subequations}
where the numerical fluxes $\mathcal{F}_{i+1/2} \approx g[\vec{n}](x_{i+1/2}) n_{i+1/2}$ are given by 
\begin{equation*}
    \mathcal{F}_{i+1/2} := \begin{cases}
        g[\vec{n}]_i n_i + \frac{1}{2}(g[\vec{n}]_{i+1} -g[\vec{n}]_{i})n_i + \frac{1}{2} g[\vec{n}]_i \text{mm}(n_{i+1}-n_{i},n_{i}-n_{i-1}), & i = 2,\dots,I-2 \\
        g[\vec{n}]_i n_i, & i = 0,1,I-1,I.
    \end{cases}
\end{equation*}
Here, we denote the minmod function by $\text{mm}(a,b)$ which is given by
\[\text{mm}(a,b) := \frac{\text{sgn}(a)+\text{sgn}(b)}{2}\min(|a|,|b|).\]
In the case of size-structured models, when it is assumed that $g[n](s_{\max}) n(t,s_{\max}) = 0$, then we set $\mathcal{F}_{I+1/2} = 0$.
System \eqref{Eq:FV_scheme} is then solved using a high-order explicit method; in this work, we used MATLAB's ode45 with default tolerances. All examples were simulated with $I = 30,000$.

\section{Approximation of the Total Population}\label{Sec:TotalPop}

Suppose we are given noisy time series data of the population number density in the form  
\(n_{k,j} := \varepsilon_{k,j} n^\star_{k,j},\)  
where, following the notation from Section \ref{Sec:WSINDy},  
\(n^\star_{k,j} := \frac{1}{|\Lambda_j|} \int_{\Lambda_j} n^\star(t_k, x) \, \diff{x},\)  
and the noise is modeled as multiplicative lognormal with \(\log(\varepsilon_{k,j}) \sim \mathcal{N}(0, \sigma^2)\).
Our goal is to approximate the true total population  
\[N^\star(t_k) := \int_\Omega n^\star(t_k, x) \, \diff{x}
\]  
using the noisy measurements \(n_{k,j}\).

We begin by observing that the expectation of the bias in the calculation is given by
\[
\mathbb{E}\left[\sum_{j=1}^J n_{k,j} - \sum_{j=1}^J n_{k,j}^\star\right] = \left(\exp\left(\frac{\sigma^2}{2}\right)-1\right) \sum_{j=1}^J n^\star_{k,j}.
\]  
This observation allows us to define an unbiased estimator for the total population when \(\sigma\) is known:  
\[
N_k := \frac{1}{\exp(\sigma^2/2)} \sum_{j=1}^J n_{k,j}.
\]
In practice, however, \(\sigma\) is typically unknown and must be estimated from the data. 
For this purpose, we employ a loess (locally estimated scatterplot smoothing) polynomial fit applied to the logarithm of the noisy density values. 
Specifically, we estimate \(\sigma^2\) by computing the variance of the log-transformed data around the loess estimate:  
\[
\tilde{\sigma}^2 := \frac{1}{KJ} \sum_{k=1}^K \sum_{j=1}^J \left(\log(n_{k,j}) - \widehat{\log(n_{k,j})} \right)^2,
\]  
where \(\widehat{\log(n_{k,j})}\) denotes the loess-fitted value for \(\log(n_{k,j})\).
For precise details on the Loess fitting strategy, we point the reader to \cite{Harrell2015}.

\section{Additional Plots and Figures}\label{Sec:AppendPlotsandFigs}
In this section, we include some additional plots demonstrating the algorithm's overall performance for the linear example problems.

\begin{figure}[ht]
    \centering    \includegraphics[width=1\linewidth,height = 0.3\linewidth]{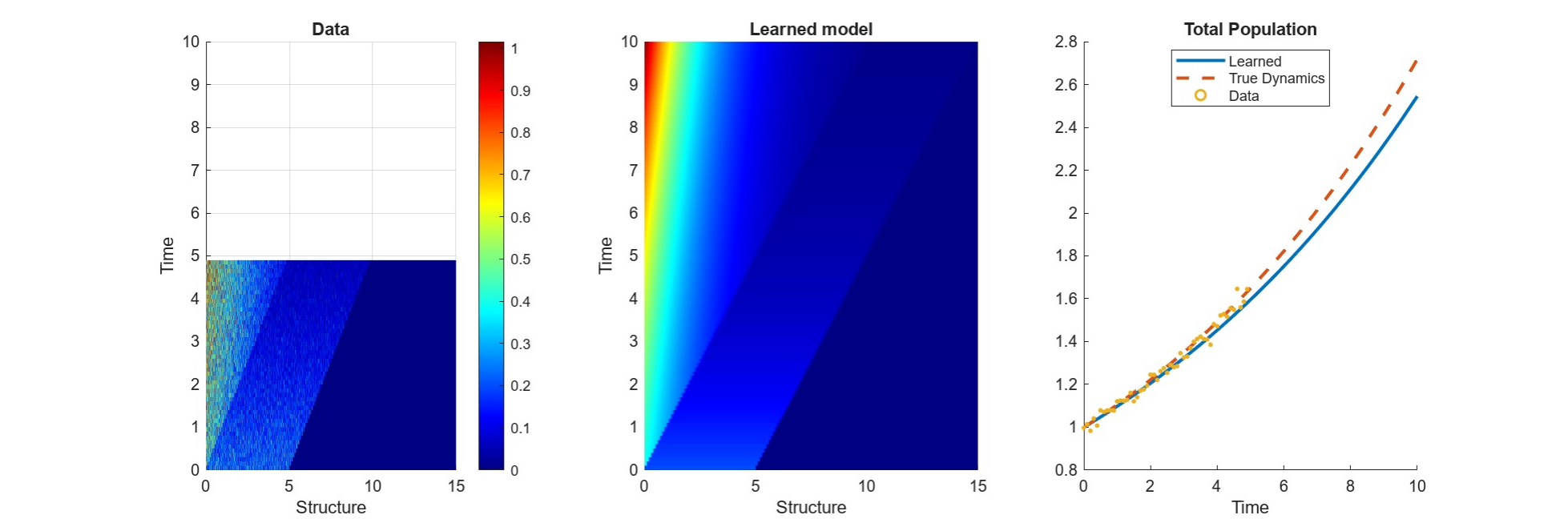}
    \includegraphics[width = 1\linewidth,height = 0.3\linewidth]{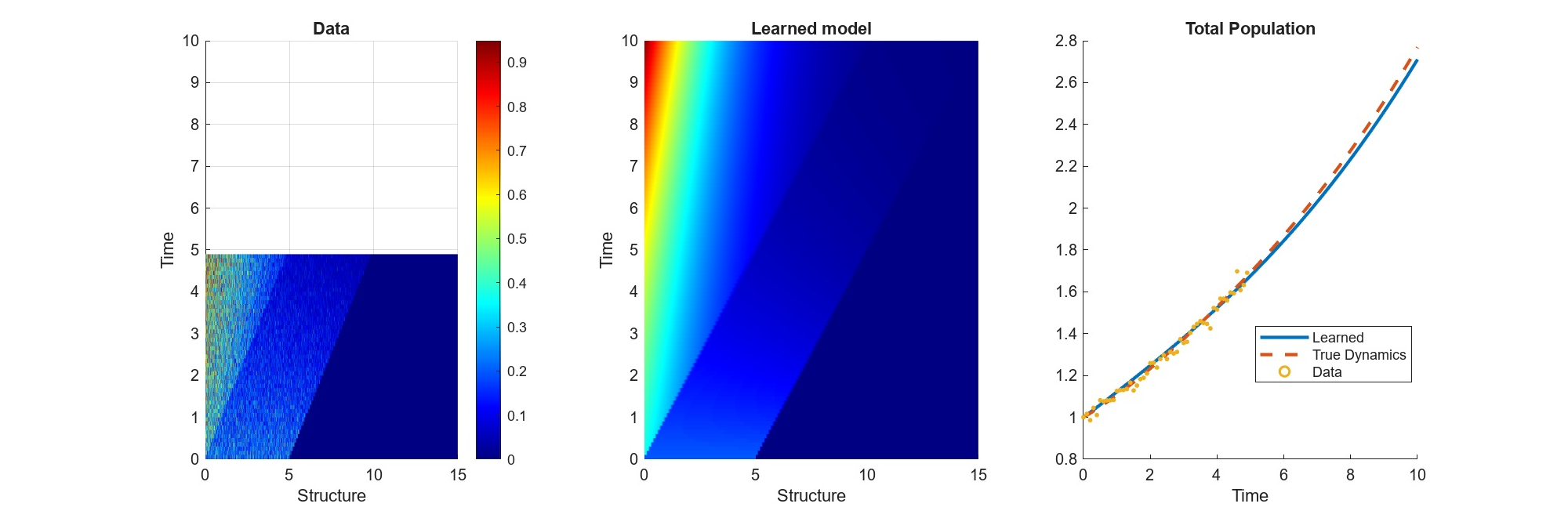}
    \includegraphics[width = 1\linewidth,height = 0.3\linewidth]{Figures/TypicalRunL3_NR066.pdf}
    \includegraphics[width = 1\linewidth,height = 0.3\linewidth]{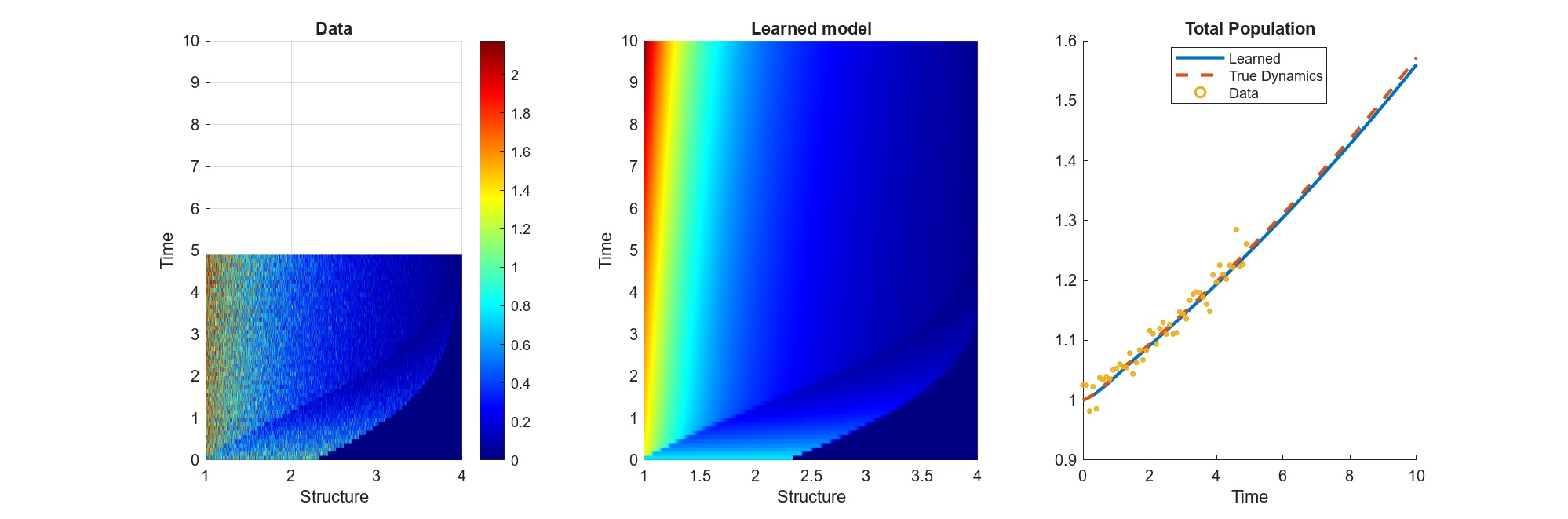}
    \caption{Typical results of using the WSINDy algorithm for the linear models in Table \ref{Table:Example problemsAge} with libraries presented in Table \ref{Table:Ex_Libraries}, $\sigma_{NR} = 0.66$, and 50 points in time. The prediction error, $\textbf{E}_p$, and true positivity ratio, \textbf{TPR}, of each learned model from top to bottom are approximately $\textbf{E}_p =$ 0.052, 0.016, 0.035, 0.013 and $\textbf{TPR}= 1, 0.6, 0.8, 1$,  respectfully.}
    \label{fig:LinearModels}
\end{figure}

\end{document}